\documentclass[amssymb,prl,aps,twocolumn,amsmath,showpacs]{revtex4-1} %,preprint
\usepackage{graphicx}
\usepackage{amsmath}
\usepackage[T1]{fontenc}
\usepackage{graphicx}
\usepackage{color}
\usepackage{times} 
\usepackage{appendix} 
\usepackage{ulem}

\newcommand{\om} {\omega_{\perp}}

\begin{document}
\title{Mapping out the quasi-condensate transition through 
the 1D-3D dimensional crossover}
\date{\today}
\author{J.~Armijo$^{(1)}$, T.~Jacqmin$^{(1)}$, K. Kheruntsyan$^{(2)}$ and I.~Bouchoule$^{(1)}$} 
\affiliation{$^{(1)}$Laboratoire Charles Fabry, UMR 8501 du CNRS,  Institut
d'Optique, 91 127 Palaiseau Cedex, France\\
$^{(2)}$ARC Centre of Excellence for Quantum-Atom Optics, School of Physical
Sciences, University of Queensland, Brisbane, Queensland 4072, Australia
}

%\email[Corresponding author: ]{julien.armijo@institutoptique.fr}

\begin{abstract}
By measuring the  density fluctuations in a highly elongated weakly interacting Bose gas, 
we observe and quantify the transition from the ideal gas to a quasi-condensate regime 
throughout the dimensional crossover from a purely 1D to an almost 3D gas. We show that 
that the entire transition region and the dimensional crossover are described surprisingly 
well by the modified Yang-Yang model. 
Furthermore, we find that at low temperatures the 
linear density at the quasi-condensate transition scales according to an 
\textit{interaction-driven} scenario of a longitudinally uniform 1D Bose
gas, whereas at high temperatures it scales according to the 
\textit{degeneracy-driven}
critical scenario of  transverse condensation of a 3D ideal gas.
\end{abstract}


%\showpacs

\maketitle

Low-dimensional (1D or 2D) systems can have physical properties dramatically 
different to their 3D counterparts. Experimental realizations of such systems 
in recent years have been particularly exciting in the field of ultracold 
atomic gases 
\cite{PetrovReview,
DalibardBloch}.
Here the reduction of dimensionality is achieved by using highly 
anisotropic trapping potentials, in which 
lowering the temperature 
leads to ``freezing'' out certain motional 
degrees of freedom to the 
respective ground state. 
 For situations when the ``freezing'' is not perfect, an intriguing
fundamental question arises: How the low-dimensional and the 
3D physics get reconciled in the 
dimensional crossover?

In this paper we address this question for a weakly interacting
Bose gas that
is confined transversely by a harmonic trap of frequency
$\omega_{\perp}/2\pi$, but is 
homogeneous and is in the thermodynamic limit
with respect to the longitudinal direction.
The 1D regime is obtained when
the thermal energy $k_BT$  and the chemical 
potential $\mu$ become much smaller than the 
transverse excitation energy $\hbar\omega_\perp$.
 In the absence of interatomic interactions, 
the homogeneous 1D gas is
characterised 
by the absence of  Bose-Einstein condensation. In the
 3D limit, however, for $k_{B}T\gg \hbar\omega_\perp$,
a sharp {\it transverse condensation} 
is expected: the atoms accumulate in the transverse ground state due to the saturation 
of population in the transversally excited states, 
yet the resulting 1D gas is still 
uncondensed with respect to the longitudinal states
~\footnote{This phenomenon was first pointed out 
by N. J. van Druten and W. Ketterle,
in Phys. Rev. Lett. \textbf{79}, 549
(1997),
for a system of finite longitudinal size, in which case the atoms 
eventually condense into the true ground state as the temperature is reduced.}.
Incorporating weak repulsive interactions, 
one expects, in the 1D limit, a 
smooth \textit{interaction-driven} transition
from the ideal gas regime towards the so-called 
quasi-condensate regime~\cite{Petrov00} 
characterized by suppressed density fluctuations
while the phase still fluctuates.
 Quasi-condensates can be also created in the 3D limit~\cite{Petrov01},
as observed experimentally~\cite{Quasi-condensate-Hannover,Quasi-condensate-Orsay}. 
  In this paper we investigate the nature of the quasi-condensate 
transition throughout the whole 1D-3D dimensional crossover.

\begin{figure*}[tbh]
\includegraphics[width=17.5 cm]{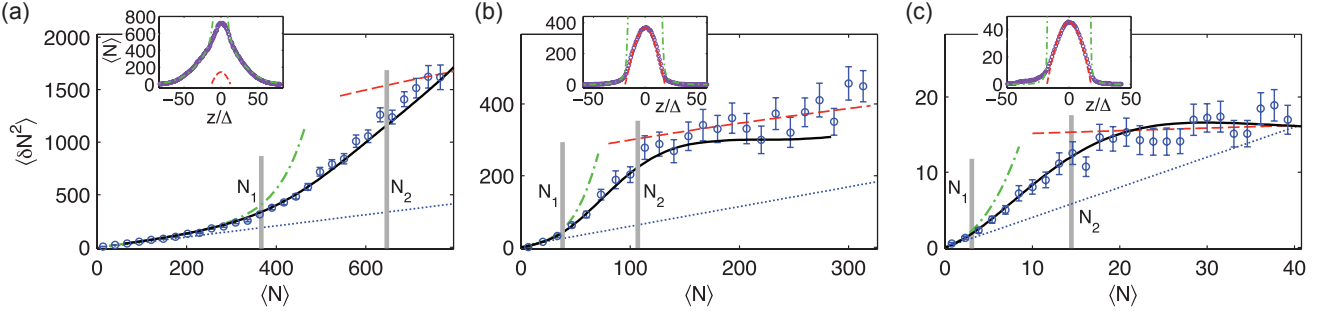}
\caption{ Density fluctuations across the quasi-condensate transition, 
for $k_BT/\hbar\omega_\perp=3.6$ (a), 1.0 (b) and 0.09 (c). 
The  measured
  atom-number variances in individual pixels, $\langle \delta N^2
  \rangle$, as a function of the mean atom number
  $\langle N \rangle$ are shown as circles, with the error bars
representing the statistical uncertainty. 
Different curves are the predictions from theoretical models, rescaled by 
the resolution factor $\kappa=0.53$, $0.55$ and $0.4$ for (a), (b) and (c), 
respectively (see text): 
 modified Yang-Yang model (solid), 
  ideal Bose gas (dash-dotted),  
 quasi-condensate (dashed) and the Poissonian shot-noise 
level (dotted). 
  The two vertical gray lines 
  give the atom numbers $N_1=\Delta n_1$ and $N_2=\Delta n_2$ (see text).
  The insets show the average density profiles, together with
  the predictions from the same models (different curves are as in the
 main graphs).
  The chemical potential $\mu_{0}$ in the trap center is
  deduced from the fit of the wings of the density profile to the
  ideal Bose gas EoS for (a),  and from the peak density and the
  quasi-condensate EoS for (b) and (c). The trap oscillation frequencies
are: $\omega_{\perp}/2\pi=3.0 $ kHz,
 $\omega_{\parallel}/2\pi=8.0 (5)$ Hz for (a-b) and
$\omega_{\perp}/2\pi=3.9$ kHz, $\omega_{\parallel}/2\pi=4.0(5)$ Hz for (c).
The absolute temperatures are $T=510$~nK (a), 160 nK (b), and 18 nK (c).
}
\label{fig.varN}
\end{figure*}

Our study relies on the measurement of atomic density fluctuations, previously used to identify the two 
limiting regimes -- the ideal gas and the quasi-condensate~\cite{Esteve06}. 
Owing to a higher measurement precision,
we now probe the transition 
itself,
 including the crossover from a deeply 1D
regime with $k_BT\!\ll \!\hbar \omega_{\perp}$ to an almost 3D
regime with $k_BT\! \simeq \!3\hbar \omega_{\perp}$. 
For our parameters, the chemical potential becomes non-negligible
compared to $\hbar\omega_\perp$ only in the quasi-condensate 
regime so that the dimensional crossover 
occuring in the quasi-condensate transition is relative to    
$T$ only.
Although
 the atoms in our experiment are trapped longitudinally, we probe 
the physics of a longitudinally {\it homogeneous} gas  
because the 
longitudinal confinement
is sufficiently weak and   the
density fluctuations are measured locally.
We find that the transition in the entire dimensional crossover 
is well described 
 by the so called  modified Yang-Yang model (MYYM)~\cite{Amerongen08}, 
in which the atoms
in the transverse ground state are treated using the 
exact
thermodynamic  solution of the
1D Bose gas model with contact
interactions \cite{YangYang69}, whereas the atoms in the transverse
excited states are treated as independent ideal 1D Bose gases.
 This
shows that the quasi-condensate transition 
maintains its 1D physical origin
even for
 temperatures  
$k_{B}T\!>\!\hbar \omega_{\perp}$.
Moreover, by monitoring the linear density at the 
quasi-condensation transition,
we show that the physics is continuously modified across the
dimensional crossover.
In the 1D regime, the transition is broad, interaction-driven, 
and 
scales as expected from 
the theory of weakly interacting gases. 
In the 3D limit, the transition is ruled
by 
the ideal gas scenario of 
 transverse condensation 
and the transition density no longer depends on the interaction strength. 
We give a theoretical prediction for the temperature of this
dimensional crossover, $T_{DC}$, and show that,
unless the interactions are extremely weak,
$k_BT_{DC}\!\sim \!\hbar \omega_{\perp}$.
Dimensional crossover in 
other ultracold gas systems has been 
studied 
both theoretically~\cite{Das02HardCoreBosons1D-3Dcrossover,Astrakharchik02QMonteCarlo1D-3D,Cazalilla04,AlKhawaja03dim-Tcrossover} 
and experimentally~\cite{Gorlitz01,Stoferle04Mot1D-3D},
but without
comparison 
continuously
throughout the crossover.

We conduct our experiment on an atom chip using $^{87}$Rb atoms in the
$\vert F\!=\!2,m_F\! = \!2 \rangle$ hyperfine state.
The on-chip current carrying micro-wires realise an Ioffe magnetic trap
 with a longitudinal oscillation frequency 
ranging from 5.0 Hz to 8 Hz and a transverse oscillation 
frequency  ranging from 3 kHz to
4 kHz.
An ultra-cold gas in thermal equilibrium is prepared using rf forced 
evaporation.
We then take \textit{in situ} absorption images of the atomic cloud using a
nearly resonant laser at $\lambda= 780$nm, as detailed
in~\cite{ArmijoSkew}. The imaging spatial resolution 
in the object plane
has an rms width of about
2~$\mu$m, whereas the camera pixel size is
$\Delta=4.5\mu$m. By summing over the transverse pixels, we derive
from the images the longitudinal atomic density profile, thus reducing
the notion of a pixel to a segment of length $\Delta$. 
The absolute
calibration of the density profiles is described
in~\cite{ArmijoSkew}. 
We perform a statistical
analysis of hundreds of images, taken under the same
experimental conditions: for each density profile and pixel, we extract
the atom number fluctuation $\delta N=N-\langle N \rangle$, 
where $N$ is the measured number of atoms in the pixel and $\langle
N\rangle$ its mean value. The fluctuations are binned according to
$\langle N\rangle$ and the variance $\langle \delta N^2 \rangle$ is
computed for each bin. 
Finally, we subtract the contribution of the
optical shot noise, which is typically less than 20\% of the atomic
fluctuations. 
Figure~\ref{fig.varN} shows typical results for $\langle
\delta N^2 \rangle$ for three different temperatures, together with
the respective average density profiles.
As the images are blurred due to finite imaging resolution, the 
measured fluctuations are reduced by a factor $\kappa$ compared to their true
values.
We deduce $\kappa$ from the measurement of atom number \textit{correlations} 
between the adjacent 
pixels, as explained  in~\cite{ArmijoSkew}.

For our experimental parameters, we can use the local density approximation
along the longitudinal dimension $z$
\cite{Kheruntsyan05}, since the
correlation length $l_c$ of density
fluctuations, the pixel length $\Delta$, 
and the cloud length 
$L$ satisfy 
$l_c\!\ll \!\Delta\!\ll \!L$.
Thus,  
 the gas contained in a pixel 
$[z,z+\Delta]$ is well described by a longitudinally
homogeneous system in the
thermodynamic limit, whose local chemical potential is 
$\mu(z)\!=\!\mu_{0}\!-\!V(z)$, where $V(z)$ is the longitudinal trapping
potential. The thermodynamic quantities 
 can be derived
from the equation of state (EoS) $n\!=\!n(\mu,T)$ for a 
longitudinally homogeneous, but transversely trapped gas,  where
$n$  
is the linear (1D) density. In particular, 
$\langle N \rangle\!=\!n\Delta$ and %n(z)=n(\mu(z),T)$, and 
the atom-number fluctuations can be calculated using the
thermodynamic relation
\begin{equation} 
\langle \delta N^2\rangle= 
k_B T \Delta (\partial  n /\partial \mu)_{T}.
\label{eq.varN}
\end{equation}

Thermometry is done in two alternative ways. For
hot gases [such as in Fig.~\ref{fig.varN}~(a)], assuming a perfectly 
harmonic longitudinal potential, 
we deduce the temperature by fitting the wings of the density profile 
to the EoS of  an ideal Bose gas,
\begin{equation}
n=\frac{1}{\lambda_{T}}\sum\nolimits_{\alpha=1}^{\infty} 
\frac{e^{\alpha \mu/k_B T}}{\sqrt{\alpha}}\frac{1}{(1-e^{-\alpha\hbar\omega_\perp/k_BT})^2}.
\label{eq.eosid}
\end{equation}
Here, $\lambda_{T}=\sqrt{2\pi\hbar^2/mk_BT}$ is the thermal de Broglie wavelength, and the EoS
is obtained summing the contributions of the 
transverse harmonic oscillator modes.

For the coldest samples, because of the lack of pixels in the ideal-gas part of the cloud, we 
deduce the temperature from the measured %atom number 
fluctuations in the quasi-condensate (central) part, 
using Eq.~(\ref{eq.varN})
and the 
quasi-condensate EoS~\cite{Fuchs03etMateo2007}
\begin{equation}
\mu=\hbar\omega_\perp \left(\sqrt{1+4na}-1\right),
\label{eq.eosqbec}
\end{equation}
valid
in the entire 1D-3D crossover region with respect to $\mu$,
where $a=5.7~$nm is the 3D scattering length.
This fluctuation-based thermometry
has an accuracy of 
about $20$\%, 
representing a viable alternative to the  thermometry based on 
the analysis of density ripples appearing after 
time-of-flight~\cite{Manz2010}.  
A related 
 fluctuation-based thermometry~\cite{Mueller10,Sanner10} 
uses the knowledge of the
longitudinal confining potential to 
deduce the gas compressibility $\partial n/\partial \mu$ 
from the density profiles. Although less general because of the assumption of 
validity of Eq.~(\ref{eq.eosqbec}),
our method has the advantage to work in not perfectly characterised 
longitudinal potentials, as is often the case in atom-chip 
experiments~\cite{Amerongen08,Esteve2004}.
%.  This is particularly valuable
%in atom-chip experiments where potential deformations
%may arise~\cite{Amerongen08} due to the micro-wire imperfections~\cite{Esteve2004}.

Once $\kappa$ and $T$ are  determined, 
the experimental data for the atom number fluctuations are compared with different theoretical models 
without any further adjustable parameters.
As we see from Fig.~\ref{fig.varN}, the two main regimes 
of a weakly interacting  
Bose
gas~\cite{KheruntsyanPRL03,Kheruntsyan05}
are clearly identified.
First, at low $\langle N\rangle$
the fluctuations follow the prediction from the ideal
gas EoS %, Eq.~
(\ref{eq.eosid}) (dash-dotted curve). 
Within this regime, but for nondegenerate samples, 
the
fluctuations are Poissonian and follow the shot-noise (dotted) line, 
as in Fig.~\ref{fig.varN}~(a) for $\langle N\rangle \!<\!200$. 
For degenerate samples (in the quantum decoherent sub-regime \cite{KheruntsyanPRL03,Kheruntsyan05}), 
 atomic bunching due to Bose statistics raises 
the fluctuations well above the shot-noise level~\cite{Esteve06,BouchouleChipBook}.
The second main regime is the quasi-condensate regime, 
where 
density fluctuations are 
suppressed by the repulsive interactions.
The 
data in Fig.~\ref{fig.varN} indeed converge  
at large $\langle N\rangle$ 
towards the prediction of the quasi-condensate EoS~(\ref{eq.eosqbec})
(dashed lines).

To describe the transition between the two main regimes,
we use the modified Yang-Yang model~\cite{Amerongen08}, whose EoS is
\begin{equation}
n=n_{YY}(\mu,T)+\sum\nolimits_{j=1}^{\infty}(j+1)n_{\rm e}(\mu_j,T).
\label{eq.eosyy}
\end{equation}
Here, the first term describes the atoms in the transverse ground state treated within the exact thermodynamic 
solution of the 1D Bose gas model \cite{YangYang69}, 
while the second term describes the atoms in the transverse excited states,
 each 
treated as an ideal Bose gas with a shifted chemical potential 
$\mu_j=\mu-j\hbar\omega_\perp$ and a linear density
$n_{\rm e}(\mu_j,T)=g_{1/2}\left(e^{\mu_j/k_BT}\right)/\lambda_T$, where
$g_{1/2}$ is a Bose function.
Since $a\ll l_\perp$ in our experiment, where $l_\perp=\sqrt{\hbar/m\omega_\perp}$ 
is the 
transverse oscillator length, we use $g=2\hbar\omega_\perp a$~\cite{Olshanii98}
as the effective 1D coupling in the MYYM.

The  transition 
 to the quasi-condensate  state 
in a 1D gas  occurs when the chemical 
potential crosses zero~\cite{Bouchoule07}, over a width 
$\mu_{t}\!=\!(mg^2/\hbar^2)^{1/3}(k_{B}T)^{2/3}$~\cite{BouchouleChipBook}.
Neglecting
correlations between the different transverse states, 
one can expect the excited state 1D gases to remain 
nearly ideal and hence the  MYYM
to correctly describe the quasi-condensate 
transition 
as long as $\mu_{t}\!\ll\!\hbar\omega_\perp$, or  $\mu_{t}/\hbar\omega_\perp\!=
\!\left[(k_BT/\hbar\omega_\perp)(a/l_\perp)\right]^{2/3}\!\ll \!1$.
Since  $a/l_\perp\!\simeq 0.03$ in our experiment, 
the MYYM can be expected to be valid up to 
temperatures 
significantly larger than 
$\hbar\omega_\perp/k_B$.
The experimental data in Fig.~\ref{fig.varN}
are indeed in remarkable 
agreement with the MYYM prediction in the entire transition region and for 
all explored temperatures.

In the quasi-condensate regime, however, 
the MYYM
underestimates the fluctuations at high densities.
Indeed, 
when $\mu$
is no longer negligible compared to $\hbar\omega_\perp$, 
the repulsive interactions produce transverse swelling of the density profile
-- an effect not taken into account in the MYYM.
This effect, which is a manifestation of the dimensional crossover 
with respect to $\mu$~\cite{Gorlitz01},
is, on the other hand, captured by the EoS~(\ref{eq.eosqbec}),
which better describes the quasi-condensate 
regime [see Fig.~\ref{fig.varN}~(b)].

\begin{figure}[tb]
\includegraphics[width=7.85cm]{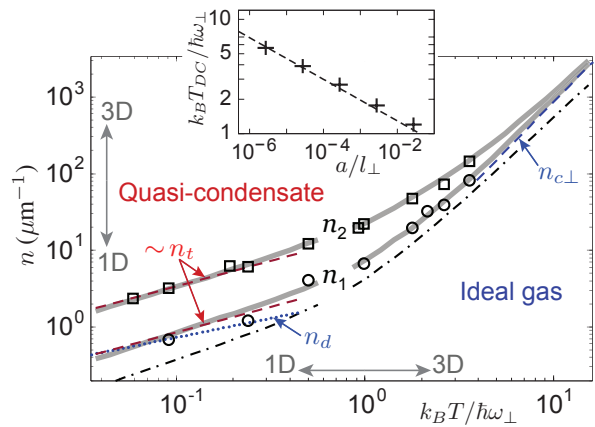}
\caption{
Phase diagram  
in the 1D-3D crossover.
The data at $k_BT/\hbar\omega_\perp\!<\!0.8$ ($>\!0.8$~kHz)
are taken with $\omega_\perp/2\pi=3.9$~kHz ($=\!3.0$~kHz). 
The measured transition boundaries $n_1$ (circles) and $n_2$ 
(squares) 
are plotted along with the predictions from 
the MYYM (solid gray curves).
The dash-dotted curve is the perturbative calculation of $n_1$;  
all other lines 
are as labeled (see text).
The inset shows the dimensional crossover temperature $T_{DC}$ versus 
$a/l_{\perp}$ (crosses), fitted 
 with a power law $(a/l_\perp)^{-2/11}$  (dashed line, see text). 
}
\label{fig.width}
\end{figure}

To quantify the quasi-condensate transition
we define the linear densities $n_1$ and $n_2$ 
for which the 
measured fluctuations are 20\% lower than the predictions 
of Eqs. (\ref{eq.eosid}) and (\ref{eq.eosqbec}), respectively. 
Plotting $n_1$ and $n_2$ against 
$k_{B}T/\hbar\omega_{\perp}$ (see Fig.~\ref{fig.width}) 
maps out the phase diagram and reveals the dimensional crossover
as we now explain.

In the 1D limit, $k_BT/\hbar\omega_\perp\!\ll \!1 $, the 
quasi-condensate transition is expected to occur 
for a degenerate gas 
around the density~\cite{KheruntsyanPRL03,Bouchoule07}
\begin{equation}
n_{t} \simeq \left[m(k_BT )^2/\hbar^2g \right]^{1/3}.
\label{eq.nco}
\end{equation}
This estimate can be obtained by considering the EoS of a highly degenerate 
ideal Bose gas
$n\!\simeq \!\sqrt{m(k_BT)^2/2\hbar^2|\mu|}$
\footnote{
Eq.~(\ref{eq.nco}) and the weak interaction condition 
$\gamma\!=\! mg/\hbar^2 n_t\ll1$ give
$(mT/\hbar^2n_t^2)\!\ll \!1$, confirming
that the gas is degenerate.},
and requiring that  $|\mu|$ becomes of the order of the interaction 
energy $gn$.
In the low-temperature (1D) part of the phase diagram,
we have fitted the rhs of Eq.~(\ref{eq.nco}) to both the
$n_1$ and $n_2$ curves, with two different prefactors $\alpha\!=\!0.28$ and $1.1$, respectively.  
As we see, the experimental data and the %predictions of the 
MYYM follow 
the scaling law of Eq.~(\ref{eq.nco}) quite well
 in this 
 part of the diagram.
 In contrast, the 
scaling law of the 1D degeneracy condition,
$n_d\!=\!1/\lambda_{T}$, does not account for the observed data, 
which implies that the quasi-condensate transition 
is governed by interactions and not by degeneracy.
Note that the transition begins for a gas that is not highly degenerate,
which is a sign that the data are  lying close to the crossover towards the 
strongly interacting regime and which explains why the quantum decoherent sub-regime barely exists in 
 Figs.~\ref{fig.varN}~(b)-(c).

In the  3D limit, $k_BT/\hbar\omega_\perp\!\gg \!1 $, $n_1$ 
converges towards the linear density $n_{c\perp}$ corresponding to the 
the ideal gas scenario of {\it transverse condensation}.
This
occurs when the peak 3D density $\rho_0$ reaches the threshold
$\rho_0\!=\!2.612../\lambda_{T}^3$, 
giving
\footnote{
$n_{c\perp}$ is 
obtained from Eq.~(\ref{eq.eosid}) 
by taking the limit
$k_{B}T/\hbar
\omega_{\perp}\!\!\rightarrow \!\!\infty$ and then $\mu\!\rightarrow \!0$.}
\begin{equation}
n_{c\perp}= g_{5/2}(1)(k_{B}T/\hbar \omega_{\perp})^{2}/\lambda_{T},
\label{eq.ncperp}
\end{equation}
where $g_{5/2}(1)\!=\!1.34..$. For linear densities higher than
$n_{c\perp}$, atoms accumulate in the transverse ground state,
although no  single quantum state is macroscopically occupied.  
The 3D
interaction parameter  at the
onset of condensation for $T<1~\mu$K
is $\rho_0 a^3\lesssim 10^{-4}$, so that
interactions have a negligible effect in this transition.  On the
other hand, the ratio $n_{c\perp}/n_{t}$ is of the order of
$[(k_BT/\hbar\omega_\perp)(a/l_\perp)^{2/11}]^{11/6}$.  For our
experimental parameters, $(l_\perp/a)^{2/11}\simeq 1.9$ so
that $n_{c\perp}/n_{t}\!\gg \!1$ as soon as $k_B T /\hbar\omega_\perp
\!\gtrsim \!3$. Thus, one expects a quasi-condensate in the
transverse ground state to emerge immediately after the transverse
condensation.
As we see from Fig.~\ref{fig.width}, Eq.~(\ref{eq.ncperp}) and 
the MYYM prediction for $n_1$ are
indeed in very good agreement with each other 
at high temperatures.

The transition width $(n_2-n_1)/2(n_1+n_2)$ is 0.6 in the 1D regime and 
decreases as the gas becomes more 3D.
Deep in the 3D regime, $n_2$ lacks, however, 
physical meaning.
%deep in the 3D regime. 
As an example,
for our experimental parameters and for
$k_{B}T=10\hbar\omega_\perp$, one expects only a 
small fraction of the 
atoms to be in the quasi-condensate transverse mode at 
linear  density $n_2$ and one expects the fluctuations to
actually exceed the quasi-condensate prediction at higher 
densities \footnote{For smaller $a$ and/or higher $T$, 
one can even have $n_2<n_1$.}.
We also note that in the 1D limit, 
%a perturbative treatment within the 
%quasi-condensate regime
the Bogoliubov theory within the quasi-condensate regime~\cite{Mora03} predicts that the 
fluctuations are \textit{increased} 
slightly when the density
is \textit{decreased} -- a feature seen in the MYYM prediction 
[see Fig.~\ref{fig.varN}~(c)], but which is not resolved experimentally.

In Fig.~\ref{fig.width}, the change of the scaling of $n_1$ 
from $T^{2/3}$ [Eq.~(\ref{eq.nco})]  to 
$T^{5/2}$ [Eq.~(\ref{eq.ncperp})]
clearly reveals the dimensional crossover. 
This phase diagram depends, however, 
on the strength of interactions through the scattering length $a$.
To investigate this  dependence, 
we compute $n_1$ as a function of $k_BT/\hbar\omega_\perp$, for several
values of
$a/l_\perp$, using the standard perturbation theory 
with respect to the 3D coupling $g_{3D}=4\pi\hbar^2a/m$,
which correctly describes departures from the ideal gas regime.
For the parameters of our experiment, with $a/l_\perp\!\simeq 0.03$, 
this calculation (dash-dotted curve in Fig.~\ref{fig.width}) 
is in qualitative agreement with the scaling from the 3D to the 1D regime.
The disagreement with the data
is mainly due to the fact that for such a (large) value of  $a/l_\perp$, 
interactions are not negligible even at linear densities smaller than 
$n_1$~\footnote{For smaller  values of $a/l_{\perp}$, 
the perturbative calculation shows much better agreement with both 
the scaling law of $n_t$ in the 1D regime and with $n_{c\perp}$ in the 3D regime.}.
If $a/l_\perp$ is decreased, the crossover towards the 3D behavior 
takes place at a higher temperature. More precisely, $n_1$ converges 
towards $n_{c\perp}$ when $n_{c\perp}\! \gg \!n_t$, i.e. 
when $k_B T /\hbar \om\! \gg\! (a/l_\perp)^{-2/11}$.
 By fitting, for each $a/l_\perp$, the 1D and 
 3D asymptotic behavior with the
 scaling laws of Eqs.~(\ref{eq.nco})
 and~(\ref{eq.ncperp}), respectively, we define 
the temperature of the 
dimensional crossover
$T_{DC}$ 
 as the point where these asymptotes intersect. 
The inset of Fig.~\ref{fig.width} shows that 
$T_{DC}$ scales as $(a/l_\perp)^{-2/11}$ as expected. We also see that 
$k_BT_{DC}$ becomes significantly larger than 
$\hbar\omega_\perp$ only for extremely small values of $a/l_\perp$.
In most experimental situations, however, 
$k_{B}T_{DC}\!\sim \!\hbar\omega_\perp$,
so that the transverse condensation 
leads immediately to the formation of a quasi-condensate.

In conclusion, we have mapped out the quasi-condensate transition 
throughout 
the  1D-3D
dimensional crossover, for 
$k_BT/\hbar\omega_\perp$ ranging from $0.06$ 
to $3.6$.
We have found that, whereas the transition is always governed by the 
1D physics, it is activated by the degeneracy-driven transverse condensation
in the 3D regime while it is interaction driven in the 1D regime.
An extension of this work would be to perform 
similar measurements in 2D gases, characterising the 2D-3D crossover
and investigating the breakdown of the 
scale invariance~\cite{Rath2010,ScaleInvariance2}. 
For 1D gases, such measurements could also
be used to investigate the crossover between the weakly and strongly 
interacting regimes. 
More generally, this work shows the power of fluctuation 
measurements as a test-bed for 
competing theoretical models for the thermodynamic
equation of state of a given physical system.

\begin{acknowledgments}
This work was supported by
the IFRAF Institute, the ANR grant ANR-08-BLAN-0165-03 and by the Australian Research Council.
\end{acknowledgments}

%%\bibliographystyle{prsty}
%\bibliography{crossoverbib}

\end{document}